\documentstyle[sprocl,epsfig]{article}

\def\nostrocostrutto#1\over#2{\mathrel{\mathop{\kern 0pt \rlap 
  {\raise.2ex\hbox{$#1$}}}
  \lower.9ex\hbox{\kern-.190em $#2$}}}
\def\gsim{\nostrocostrutto > \over \sim}   

\newcommand{\be}{\begin{equation}}
\newcommand{\ee}{\end{equation}}
\newcommand{\ba}{\begin{eqnarray}}
\newcommand{\ea}{\end{eqnarray}}

\newcommand{\eref}[1]{(\ref{#1})}      

\catcode`@=11
\newcount\@tempcntc
\def\@citex[#1]#2{\if@filesw\immediate\write\@auxout{\string\citation{#2}}\fi
  \@tempcnta\z@\@tempcntb\m@ne\def\@citea{}\@cite{\@for\@citeb:=#2\do
    {\@ifundefined
       {b@\@citeb}{\@citeo\@tempcntb\m@ne\@citea\def\@citea{,}{\bf ?}\@warning
       {Citation `\@citeb' on page \thepage \space undefined}}%
    {\setbox\z@\hbox{\global\@tempcntc0\csname b@\@citeb\endcsname\relax}%
     \ifnum\@tempcntc=\z@ \@citeo\@tempcntb\m@ne
       \@citea\def\@citea{,}\hbox{\csname b@\@citeb\endcsname}%
     \else
      \advance\@tempcntb\@ne
      \ifnum\@tempcntb=\@tempcntc
      \else\advance\@tempcntb\m@ne\@citeo
      \@tempcnta\@tempcntc\@tempcntb\@tempcntc\fi\fi}}\@citeo}{#1}}
\def\@citeo{\ifnum\@tempcnta>\@tempcntb\else\@citea\def\@citea{,}%
  \ifnum\@tempcnta=\@tempcntb\the\@tempcnta\else
   {\advance\@tempcnta\@ne\ifnum\@tempcnta=\@tempcntb \else \def\@citea{--}\fi
    \advance\@tempcnta\m@ne\the\@tempcnta\@citea\the\@tempcntb}\fi\fi}
\catcode`@=12

\begin{document}

\setcounter{page}{0}
\thispagestyle{empty}

\noindent
\rightline{MPI-PhT/97-3}
\rightline{January 1997}
\vspace{1.0cm}
\begin{center}
{\Large \bf Soft particle production in QCD jets} 
\end{center}
\vspace{0.4cm} 
\begin{center}
SERGIO  LUPIA~\footnote{E-mail: lupia@mppmu.mpg.de}  
\end{center}

\begin{center}
\mbox{ }\\
 {\it Max-Planck-Institut f\"ur Physik \\
(Werner-Heisenberg-Institut) \\
F\"ohringer Ring 6, D-80805 Munich, Germany} 
\end{center}
\vspace{1.0cm}

\begin{abstract}
The perturbative QCD approach, based on Modified Leading Log
Approximation and Local Parton Hadron Duality, is shown to describe 
inclusive features of multiparticle production in the soft region. 
Further predictions of this approach are discussed.
\end{abstract}

\vfill 
\noindent 
to appear in the Proc. of the 33rd Eloisatron Workshop 
``Universality features in multihadron production and the leading effect'', 
(Erice, Italy, 19-25 October 1996) 
\vfill 

\newpage

\title{SOFT PARTICLE PRODUCTION IN QCD JETS} 

\author{SERGIO LUPIA} 

\address{Max-Planck-Institut f\"ur Physik (Werner-Heisenberg-Institut) \\
F\"ohringer Ring 6, D-80805 Munich, Germany\\ 
E-mail: lupia@mppmu.mpg.de} 

\maketitle
\abstracts{
The perturbative QCD approach, based on Modified Leading Log
Approximation and Local Parton Hadron Duality, is shown to describe 
inclusive features of multiparticle production in the soft region. 
Further predictions of this approach are discussed.}

\section{Introduction}

The study of soft particle production is of particular interest since 
new information  on the physics of confinement can be obtained.  
An important  open question in this subject is 
 whether the soft momentum region is smoothly connected to the hard region,
 where the purely perturbative description is appropriate, 
 or dramatic changes are present due to the dominance of the 
 hadronization process. 
In this respect, it is worth noticing that the charged particle 
inclusive energy spectra measured in $e^+e^-$ annihilation are surprisingly
close over the whole momentum range (down to small momenta of a few hundred
MeV) to the predictions of the perturbative QCD approach, based on the Modified Leading Log
Approximation (MLLA)\cite{bcm,dkmt1} and the notion of Local Parton Hadron
Duality (LPHD)\cite{ahm1,ko} (see also \cite{khozeerice}). 

In this paper I discuss: {\it a)} how to test  the validity of 
the perturbative description in the soft region and {\it b)} 
how to study the sensitivity  of experimental data to
different aspects of this approach, in particular to coherence effects 
(see \cite{ochserice} for a discussion of the sensitivity of soft particle
production to the running of the QCD coupling). 
I focus in particular on 
the study of the invariant density $E dn/d^3p \equiv dn/dy d^2p_T$ in the limit
of vanishing rapidity $y$ and transverse momentum $p_T$ or, equivalently, for
vanishing momentum $|\vec p | \equiv p$, i.e., 
\be
I_0 = \lim_{y \to 0, p_T \to 0} E \frac{dn}{d^3p} = 
\frac{1}{2} \lim_{p \to 0} E \frac{dn}{d^3p} 
\label{izero}
\ee
where the factor 1/2 takes into account that both hemisphere are added in the
limit $p \to$ 0. 

The invariant hadronic density in $e^+e^-$ annihilation 
is shown to approach a $cms$-energy independent value at low particles' momentum, 
in agreement with the predictions of the perturbative approach with the
inclusion of coherence. 
This result would then suggest to 
extend the validity of the theoretical picture down to the very soft region,
thus lending new support to the picture of LPHD,  and it
would confirm the relevance of coherence effects in soft particle production. 
To establish this interpretation more definitely, 
further tests of the perturbative picture are proposed. In particular, 
new predictions for the yield of the soft radiation in 3-jet events in
$e^+e^-$ annihilation are presented.  
A test of universality of soft particle production in different reactions
corresponding to different partonic colour emitters is also discussed. 

The results contained in this paper have been obtained in collaboration with
Valery A. Khoze and Wolfgang Ochs\cite{klo2}
 and have been partially presented elsewhere\cite{nijmegen,conf}.

\section{The invariant density $E \frac{dn}{d^3p}$ in $e^+e^-$ annihilation}

From the theoretical point of view, 
the coherence of the gluon radiation requires that 
very soft partons (with long wave length) 
 are emitted by the total colour current, totally independently 
of the internal structure of the jet. According to the hypothesis of LPHD, 
one would then expect 
that the hadron spectrum at low momentum $p$ should be 
nearly independent of the jet energy $E_{jet}$, i.e., of the $cms$ 
energy\cite{adkt1,vakcar}. 
The analysis of the inclusive spectra in the variable $\log p$,  
as measured by the  TASSO \cite{tasso} and 
 OPAL Collaborations\cite{opal1}, suggests indeed that the soft tail of the
 spectrum is approximately independent of the $cms$ energy in agreement with
 LPHD expectations. 
  In order to characterize better this phenomenon, it has recently been
 proposed\cite{lo} to study  the 
invariant density $E dn/d^3p$ at very low particle energies of the order 
of few hundred MeV. The important prediction of the perturbative QCD approach 
is that the invariant density approaches in this limit a value independent of
$cms$ energy. 
In the following, experimental analyses and theoretical
predictions for the invariant density in the soft region 
are presented in more details.

\subsection{Experimental results for low momentum particles}

\begin{figure}
\vfill \begin{minipage}{.45\linewidth}
          \begin{center}
\mbox{\epsfig{file=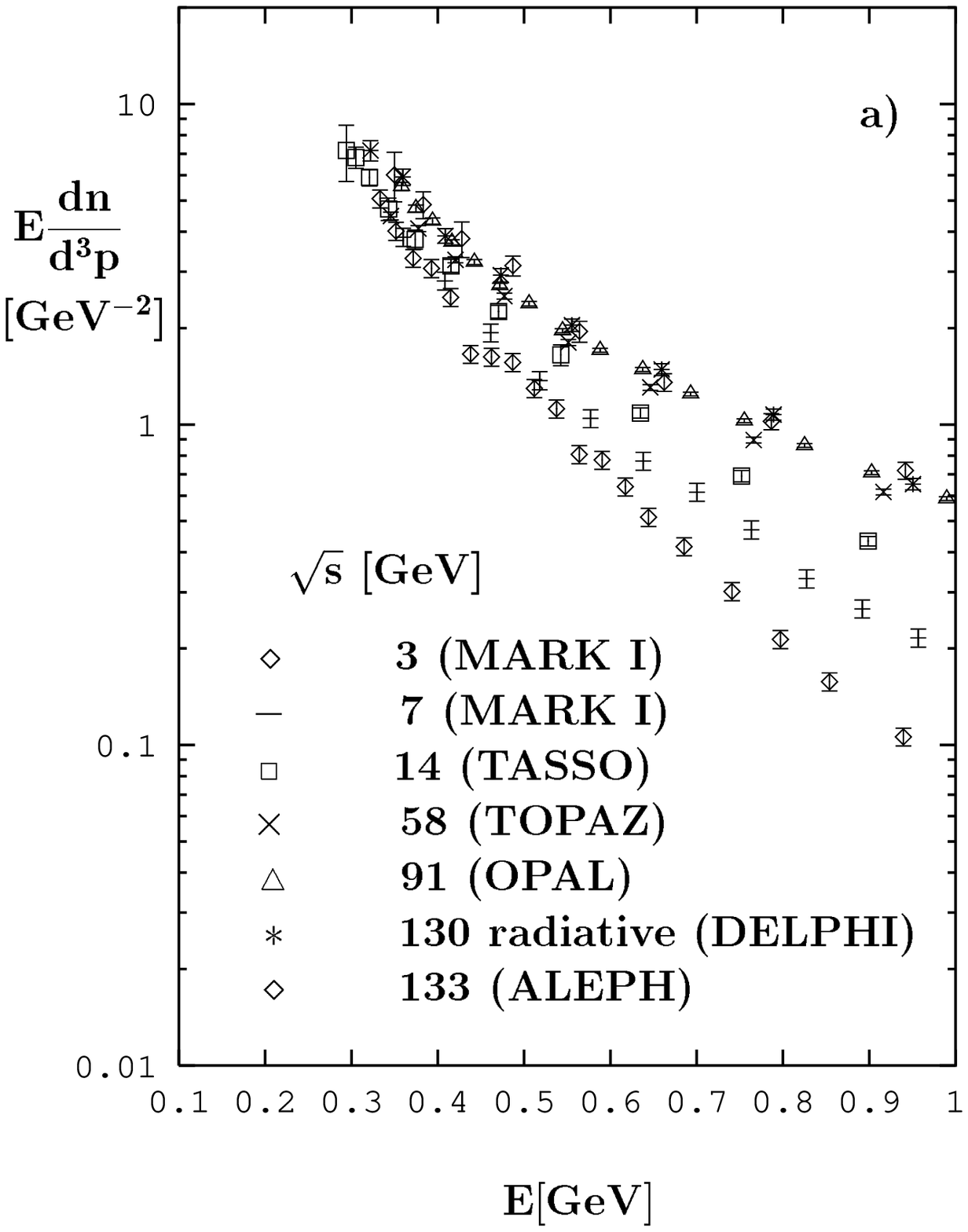,width=.6\linewidth,bbllx=5.5cm,bblly=10.cm,bburx=13.5cm,bbury=27.5cm}}
          \end{center}
      \end{minipage}\hfill
      \begin{minipage}{.45\linewidth}
          \begin{center}
\mbox{\epsfig{file=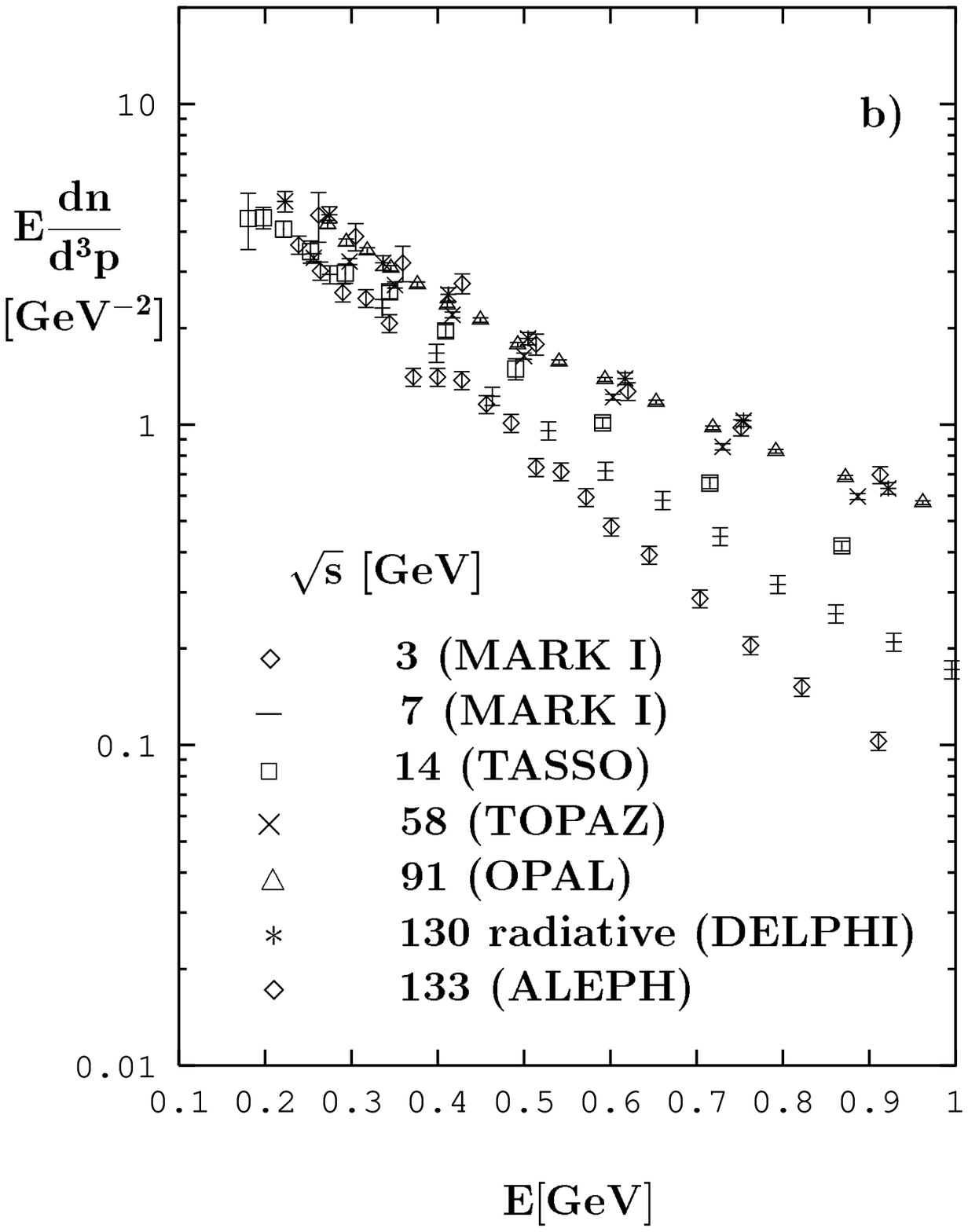,width=.6\linewidth,bbllx=5.5cm,bblly=10.cm,bburx=13.5cm,bbury=27.5cm}}
          \end{center}
      \end{minipage}
\caption{{\bf a)} Invariant density $E dn/d^3p$ of charged particles 
in $e^+e^-$ annihilation 
as a function of the particle energy $E = \protect\sqrt{p^2 + m_h^2}$ 
at $m_h$ = 270 MeV at various $cms$ energies;  
{\bf b)} the same as in {\bf a)}, but at $m_h$ = 138 MeV.}
\label{data}
\end{figure}

Let us first consider the invariant density $E dn/d^3p$ 
for all charged particles. The angular dependence is actually integrated out 
in the following, so let us call $E dn/d^3p$ the quantity 
$\frac{E}{4 \pi p^2} \frac{dn}{dp}$. 
 Since no direct measurements of this observable 
have been performed, experimental information can be extracted 
from the published data on the inclusive momentum spectrum, $dn/d\xi_p$ vs.
$\xi_p$, where $\xi_p \equiv  \log  (1/x_p) = \log (\sqrt{s}/2p)$, using 
 the relation 
\be
E \frac{dn}{d^3p} = \frac{E}{4 \pi p^3} \frac{dn}{d\xi_p}  
\label{datapre}
\ee 
where $E^2 = p^2 + m_h^2$ and $m_h$ is an effective particle mass. 

There is not a unique prescription for the choice of this effective mass; 
on one hand, the value of $m_h$ = 270 MeV has been used in
\cite{lo} to give a good description of the moments of the
energy spectra with the MLLA  formulae. 
On the other hand, since pions dominate in the soft region, the pion mass would
be a natural value for this parameter. 

Fig.~\eref{data}  shows 
the charged particle  invariant density in $e^+e^-$ annihilation 
as a function of the particle energy $E$ at different $cms$ energies 
ranging from 3 GeV up to LEP-1.5 $cms$ 
energy (133 GeV)\cite{tasso,opal1,data}; two different values of $m_h$ 
are used. 
It is remarkable that the data from all  $cms$ energies tend to converge in the
soft limit. The choice of $m_h$ does not modify the gross features of the
invariant density, although two different values of the soft limit $I_0$ 
are observed. More precisely, let us point out that the 
LEP data seem to tend to a limiting value larger by about 20\% than the data at
 lower $cms$ energies. This may be due to 
the overall  systematic effect in the relative normalization 
of the different experiments (in this
respect, let us stress that the invariant density is to a better
approximation energy independent within sets of data collected at 
the same detector). 
Alternatively, a possible physical source of energy dependence is given 
by particles produced via weak decays. They should indeed be added 
incoherently to the particles produced from the primary quarks and thereby
could yield a rise of the soft particle spectrum with increasing energy.

\begin{figure}
\vfill \begin{minipage}{.45\linewidth}
          \begin{center}
\mbox{\epsfig{file=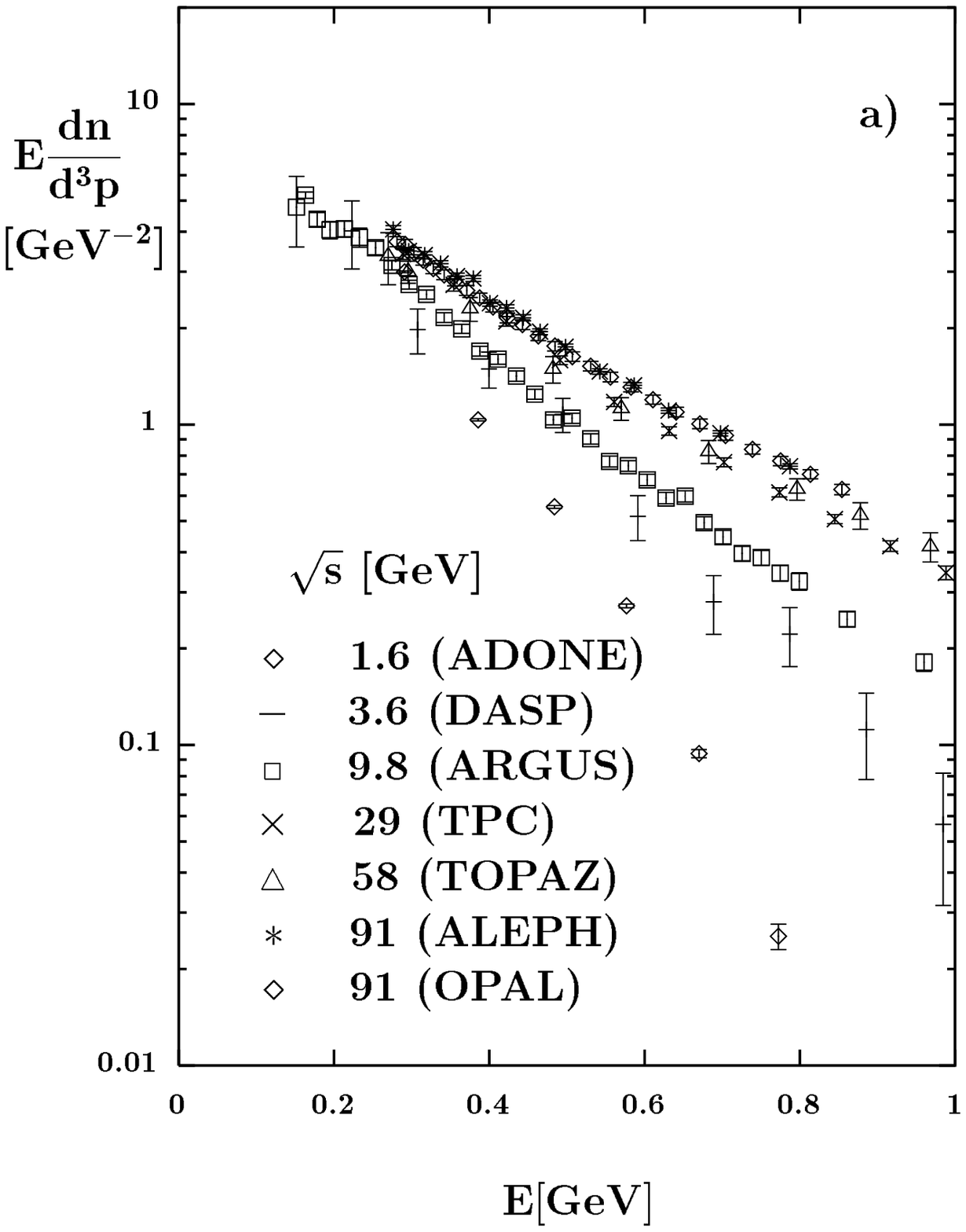,width=.6\linewidth,bbllx=5.5cm,bblly=10.cm,bburx=13.5cm,bbury=27.5cm}}
          \end{center}
      \end{minipage}\hfill
      \begin{minipage}{.45\linewidth}
          \begin{center}
\mbox{\epsfig{file=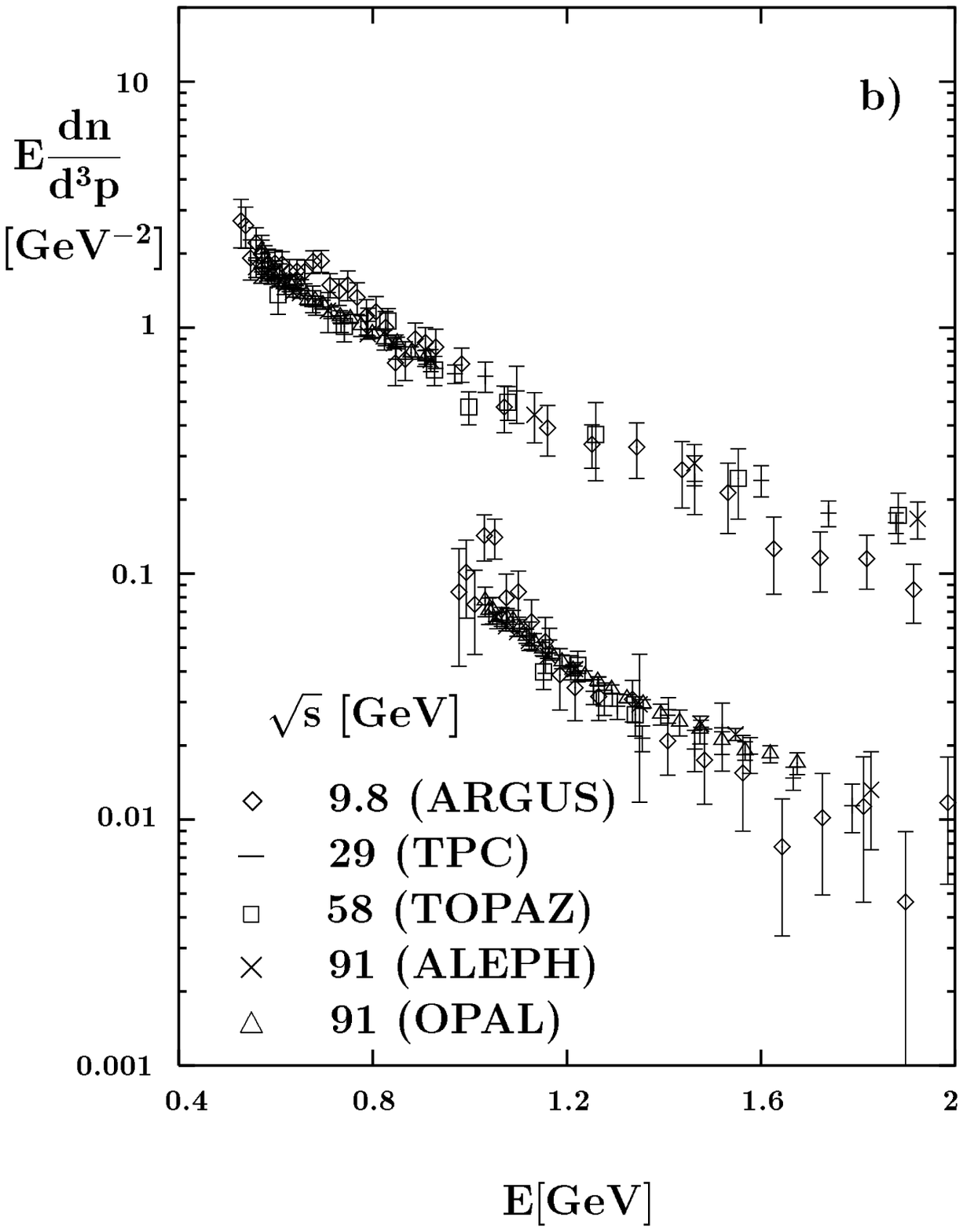,width=.6\linewidth,bbllx=5.5cm,bblly=10.cm,bburx=13.5cm,bbury=27.5cm}}
          \end{center}
      \end{minipage}
\caption{{\bf a)} Invariant density $E dn/d^3p$ of charged pions in $e^+e^-$
annihilation 
as a function of the particle energy $E$ at 
various $cms$ energies;
{\bf b)} the same as in {\bf a)}, but for charged kaons  and protons 
\protect\cite{data,datapion}. The kaon 
distribution is multiplied by a
factor 10 for the sake of clarity.} 
\label{dataid}
\end{figure}

Let us now turn to the invariant density for 
charged pions, charged kaons and  protons. 
Fig.~\eref{dataid} shows 
the invariant cross section $E dn/d^3p$ for $\pi$'s, $K$'s and $p$'s 
as a function of the particle energy $E$, as extracted from 
the  inclusive momentum spectra measured  at $cms$ energies from
1.6 GeV to 91 GeV\cite{data,datapion} according to eq.~\eref{datapre}. 
In this case, the simplest choice for the mass scale $m_i$ is given by the
particle mass itself. 
In all cases the data tend towards an energy-independent 
limit for $E \to m_i$ ($p \to 0$). 
Notice that the value of $I_0^i$ is roughly proportional to $m_i^{-2}$. 

\subsection{Theoretical predictions on the spectrum shape}

The description of the soft region of inclusive energy spectra requires some
care in the treatment of kinematics. 

Experimental hadronic spectra are usually presented as function of momentum 
$p$ or $\xi_p=\log (1/x_p)$, and they do not have any kinematical boundary 
from above. On the other hand, in the MLLA the partons are
treated as massless but the transverse momentum is required to be larger than 
the infrared cutoff $Q_0$,  such that $E=p \geq p_T \geq Q_0$. 
The theoretical predictions are 
then limited from above, as  $\xi\leq Y = \log \sqrt{s}/2Q_0$.  

As far as $E \simeq p \gg Q_0$, the different 
kinematical boundaries can be neglected and the analytical prediction using 
LPHD is uniquely defined. 
However, the deeper one goes into the soft region, the more sensitive one becomes
to the different kinematical thresholds.  In the soft 
region, there is not a unique procedure for deriving analytical predictions 
applying LPHD, but one has
to introduce some additional assumptions and define practical prescriptions. 
As a general rule, one may require that the invariant density 
$E dn/d^3p$ obtained at hadron level by one of these prescriptions 
 approaches a constant limit for $p\to 0$ as observed experimentally. 
In the following two possible alternative Ans\"atze are discussed. 

It is important to stress that the most important prediction of the
perturbative QCD approach, i.e., the energy independence of the invariant density 
for particles of low momentum, does not depend  on the particular 
Ansatz one uses. 
This feature will indeed be recovered in both Ans\"atze we are
about to discuss. 

\bigskip
\noindent {\it Ansatz 1} 
\medskip 

Let us relate  hadron and parton spectra in a single $A$-jet in the following
way\cite{dfk1,lo}: 
\be
E_h \frac{dn(\xi_E)}{dp_h} = K_h E_p \frac{dn(\xi_E)}{dp_p}
    \equiv K_h D_A^g(\xi_E,Q_0,\Lambda)
    \label{phrel}
\ee
with $E_h=\sqrt{p_h^2+Q_0^2}=E_p \equiv E
\geq Q_0$, $\xi_E\equiv \xi=\log Q/E$ and 
 $K_h$  a normalization
parameter. If hadrons from both hemispheres are added, $K_h$ should be replaced
by $2K_h$. 

Accordingly, the invariant density $E dn/d^3p$ to be compared with the data 
in $e^+e^-$ annihilation is given by: 
\be
E \frac{dn}{d^3p} = \frac{K_h}{4 \pi p^2} \frac{2 C_F}{N_C}  
D(\xi, Q_0, \Lambda)
\label{ansatz1}
\ee
with $E = \sqrt{s}/2 e^{-\xi}$ and $p^2 = E^2 - Q_0^2$ and $D(\xi,Q_0,\Lambda)$
is the inclusive spectrum predicted for a single gluon jet. 
 $C_F$ and $N_C$ are the colour factors for quark and gluon jets, 
 $C_q = C_F$ and $C_g = N_C$; 
 the ratio $C_F/N_C$ is needed to obtain the spectrum in a single quark-jet and
the factor 2 to take both hemispheres into account. 

Let us notice that, 
as far as the theoretical spectrum $D(\xi, Q_0, \Lambda)$ linearly 
goes to 0  for $p \to 0$, as in MLLA (see below), 
the invariant density for hadrons 
approaches indeed a finite limit as in (\ref{izero})
\be
I_0=K_h \frac{C_A\beta^2}{8\pi N_C\lambda Q_0^2}.
\label{limit}
\ee 
with $\beta^2 = 4N_C/b$, $b \equiv (11 N_C - 2 n_f)/3$, and 
$\lambda \equiv \log Q_0/\Lambda$, where $\Lambda$ is the QCD-scale; 
$N_C$ and $n_f$ are 
the number of colours and of flavours respectively. 

In addition, 
this prescription is particularly suitable for phenomenological analyses 
of the moments of the 
inclusive energy spectrum\cite{lo}. In this respect, let us notice
that a good phenomenological description of the charged particle inclusive 
spectrum has been obtained with $Q_0$ = 270 MeV; this value of $Q_0$ will then
be adopted in the theoretical calculations.

To obtain the theoretical prediction for the invariant density $E dn/d^3p$
within this scheme, let us then obtain an explicit solution for the spectrum  
$D(\xi, Q_0,\Lambda)$  in the soft limit. 
Consider first the Double Log Approximation (DLA),  in which energy
conservation is neglected and only the leading singularities in the parton
splitting functions are kept. 
The evolution equation of the inclusive energy distribution of partons $p$
originating from a primary parton $A$ is given by\cite{dfk1}: 
\be
D_A^p(\xi,Y) = \delta_A^p\delta(\xi) + \int_0^{\xi} d\xi' \int_0^{Y-\xi} 
dy' \frac{C_A}{N_C} \gamma_0^2(y'+\xi') D_g^p(\xi',y'+\xi') 
\label{evoleq}
\ee
where 
 $\xi = \log (1/x) = \log (Q/E)$ and $Y = \log (Q/Q_0)$ with
$E$ the particle energy and $Q$ the jet virtuality ($Q=P\Theta$ for
a jet of primary momentum $P$ and half opening angle $\Theta$); 
$C_A$ is the respective colour factor for quark and gluon jets; 
 $\gamma_0$ denotes the anomalous dimension of multiplicity and 
is related to the 
QCD running coupling by $\gamma_0^2 = 4 N_C \alpha_s / 2 \pi  
= \beta^2/\log(p_T/\Lambda)$.  

Eq.~\eref{evoleq} can be solved iteratively;  
with two iterations one gets: 
\ba
\label{duetermini}
D_A^g(\xi,Y) &=& \delta_A^g \delta(\xi) + \frac{C_A}{N_C} 
\beta^2 \log \left( 1 + \frac{Y-\xi}{\lambda}  \right) \\ 
&\times& \left[ 1 + \frac{\beta^2 \int_0^{Y-\xi} 
d\tau \log (1 + \frac{\tau}{\lambda}) 
\log (1 + \frac{\xi}{\tau+\lambda})}{\log (1 + \frac{Y-\xi}{\lambda})} \right] 
+ \dots \nonumber 
\ea
The term of order $\beta^2$, corresponding to a single gluon emission, 
yields the leading contribution for $E \to Q_0$. 
It is worth noting that this term does not depend on the $cms$ energy and it is
 proportional to the colour charge factor of the primary parton. 
 Therefore, we have explicitly found the features of soft radiation 
 which we have qualitatively  expected. 
In addition, the second iteration provides us with a 
energy-dependent term, which allows to describe  the rise 
of the spectrum with increasing $\sqrt{s}$ at particles' energies around 1 GeV. 
Let us also point out that the $\xi$-spectrum vanishes in the soft
limit $\xi\to Y$ ($E\to Q_0$) as  
\be 
D_A^g(\xi,Y) \sim Y- \xi \sim  \log E/Q_0 \sim E-Q_0.
\label{limitth} 
\ee

\begin{figure}
          \begin{center}
\mbox{\epsfig{file=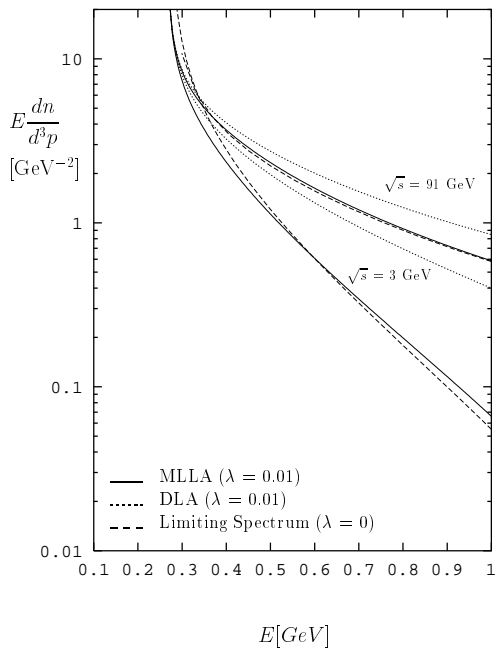,bbllx=4.5cm,bblly=9.5cm,bburx=16.5cm,bbury=28.cm,height=9cm}}
       \end{center}
\caption{Invariant density $E dn/d^3p$ 
as a function of the particle energy $E$ for $Q_0$ = 270 MeV. 
Predictions  at $cms$ energies of $\protect\sqrt{s}$
= 3 GeV (lower three curves) and 91 GeV (upper three curves)  
using eq.~\protect\eref{ansatz1} with $D_g^g$
computed in MLLA (eqs.~(\protect\ref{dmlla},\protect\ref{duetermini})), 
DLA (eq.~\protect\eref{duetermini}) 
and the Limiting  Spectrum (normalization $K_h$ = 0.45, 0.45 and 1.125
respectively).} 
\label{mllatheory}
\end{figure}

In order to get an improved description at nonasymptotic $cms$ energies, 
it is important to use the MLLA \cite{dkmt1,ahm1}, 
which takes into account the 
 exact form of the parton splitting functions
in the evolution equation.  
The MLLA solution is in general much more involved than the DLA one, but in the
soft limit the expression simplifies and one gets a 
simple exponential suppression factor:  
\begin{equation}
D(\xi,Y,\lambda)|_{MLLA} = D(\xi,Y,\lambda)|_{DLA} \exp \biggl[-a\int^Y_\xi
\gamma_0^2(y)/(4N_C) dy \biggr] 
\label{dmlla}
\end{equation} 
with $a = 11 N_C/3 + 2 n_f/3 N_C^2$. 
 This solution is valid in the soft limit
only, when the difference $[\gamma_0^2(\xi)- \gamma_0^2(Y)]$ is small
compared to $\gamma_0^2 D$. 
As a consistency check of eq.~\eref{dmlla}, 
the MLLA with fixed coupling has
been explicitly solved and eq.~\eref{dmlla} has been found 
to be  exactly satisfied in the full energy region. 

\begin{figure}
          \begin{center}
\mbox{\epsfig{file=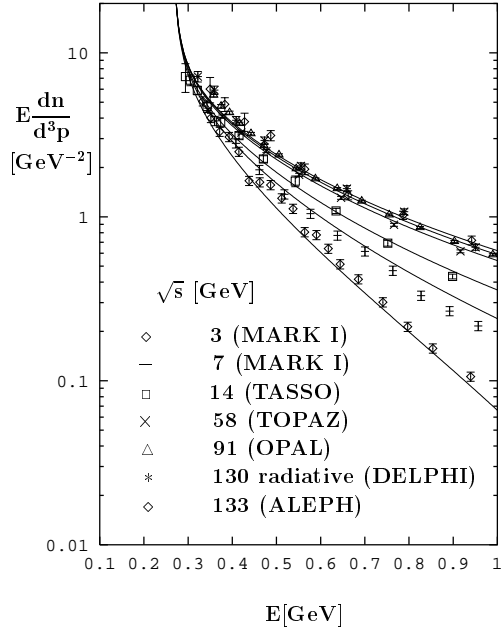,bbllx=4.5cm,bblly=9.5cm,bburx=16.5cm,bbury=28.cm,height=9cm}}
       \end{center}
\caption{Invariant density $E dn/d^3p$ of charged particles
in $e^+e^-$ annihilation 
as a function of the particle energy 
$E=\protect\sqrt{p^2+Q_0^2}$ at $Q_0$ = 270 MeV. 
Data points at various $cms$ energies
from  SLAC, TASSO and TOPAZ Collaborations, LEP-1 and
LEP-1.5\protect\cite{tasso,opal1,data} are compared to MLLA predictions 
($\lambda$ = 0.01, $K_h$ = 0.45).}
\label{chargedall}
\end{figure}

In order to estimate to what extent the iterative solution gives a good
description of the inclusive spectrum, one can consider the
inclusive spectrum $D(\xi,Q_0,\Lambda)$ in the limiting case $Q_0 =
\Lambda$. In this case, the solution, called Limiting Spectrum, is explicitly
known\cite{dkmt1} and one can avoid the iterative procedure. 
Notice that the Limiting Spectrum does not have
the same behaviour near the boundary $\xi \to Y$  as the full solution which
follows from the integral equation~\eref{evoleq} with eq.~\eref{dmlla}, 
as the Limiting
Spectrum goes to a constant as $E \to Q_0$. The invariant density computed from
the Limiting Spectrum according to eq.~\eref{ansatz1} 
shows then  a  singularity in the soft limit, $p \to 0$.

To illustrate the above analytical results, 
the predictions for the invariant density $E dn/d^3p$ in the 
low particle energy region at the two $cms$ energies of 3 and 91 GeV, 
obtained via eq.~\eref{ansatz1} in 
DLA (eq.~(\ref{duetermini})) and  MLLA (eq.~(\ref{dmlla})) with $\lambda=0.01$ 
and in the Limiting Spectrum case  are shown in Fig.~\eref{mllatheory}. 
The normalization of the limiting spectrum is as in fits to $e^+e^-$ 
annihilation; the normalizations of the
DLA and MLLA curves are chosen to approach  the
limiting spectrum for energies $E~\geq~0.5$ GeV. 

Both in DLA and in MLLA  
the invariant density  approaches an
energy independent value in the soft limit $\xi\to Y$ ($E\to Q_0$). This
originates from the soft gluon emission contribution of order $\alpha_s$ 
which is determined
by the total colour charge of the primary partons due to the colour
coherence. In this limit the MLLA converges towards the DLA. 
The MLLA solution is in good agreement with the Limiting Spectrum solution,
thus showing that the iterative solution includes indeed the dominant contribution
in the soft region.  

A quantitative comparison of the theoretical predictions of MLLA 
with experimental data for the invariant density of all charged
particles is shown in Fig.~\eref{chargedall}. 
A good agreement is visible, both in the energy independence of 
the soft limit $I_0$ and in the energy dependence of the invariant density 
at particle energies above 500 MeV. 

It is also possible to compute the  predictions of MLLA with fixed
coupling. In this case a full solution can be obtained\cite{lo,nijmegen}.  
It is interesting to notice that the model with fixed coupling cannot describe
the steep slope at very low momentum, thus suggesting that the running of the
coupling is relevant in this region. Further studies of models, where 
the coupling is frozen below a certain momentum threshold, are in progress 
(see also \cite{lo,ochserice} for further discussions of the effect of the
running of the coupling in inclusive energy spectra). 

\bigskip
\noindent {\it Ansatz 2} 
\medskip 

\begin{figure}
          \begin{center}
\mbox{\epsfig{file=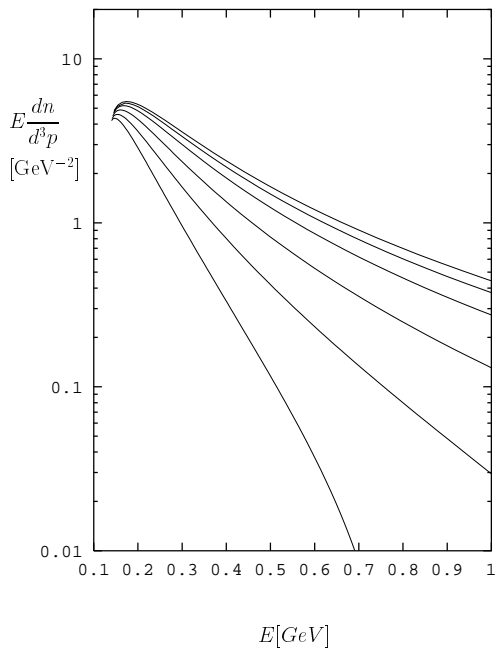,bbllx=4.5cm,bblly=9.5cm,bburx=16.5cm,bbury=28.cm,height=9cm}}
       \end{center}
\caption{Invariant density $E dn/d^3p$ 
as a function of the particle energy $E = \protect\sqrt{p^2 + Q_0^2}$ 
with $Q_0$ = 138 MeV predicted at various $cms$ energies 
using eq.~\protect\eref{phrel1} ($K_h$ = 1.125).}
\label{khozetheory}
\end{figure}

Let us consider the alternative identification\cite{dkt9}: 
\be 
  \frac{dn}{d\xi_p} = \biggl( \frac{p}{E} \biggr)^3 D_{\lim}(\xi_E,Y) 
\ee
where $D_{\lim}(\xi_E,Y)$ indicates the Limiting Spectrum solution. 
With this Ansatz, the invariant density for hadrons 
in $e^+e^-$ annihilation is given by: 
\be
E \frac{dn}{d^3p} =  \frac{K_h}{4 \pi E^2}  \frac{2 C_F}{N_C}  
D_{\lim}(\xi_E,Y)
\label{phrel1}
\ee
The preferred value for $Q_0=\Lambda$ is in this case the pion mass, 
as natural in the soft region, where most produced particles are indeed pions. 

Let us notice that as $p \to 0$, the particle energy $E$ goes to the finite
value $Q_0$ and $D_{\lim}$ goes to a constant, $D_{\lim}^0$. 
One then recovers again a finit value for the soft limit 
\be
I_0 = \frac{K_h}{4 \pi Q_0^2}  \frac{2 C_F}{N_C} D_{\lim}^0 
\ee
The invariant densities predicted with this method at different $cms$ energies 
are shown in Fig.~\eref{khozetheory}.  
It is important to stress that also with this second Ansatz 
all curves tend to converge to a common value
independent of $cms$ energy at very low particles' energies. 

\begin{figure}
\vfill \begin{minipage}{.45\linewidth}
          \begin{center}
\mbox{\epsfig{file=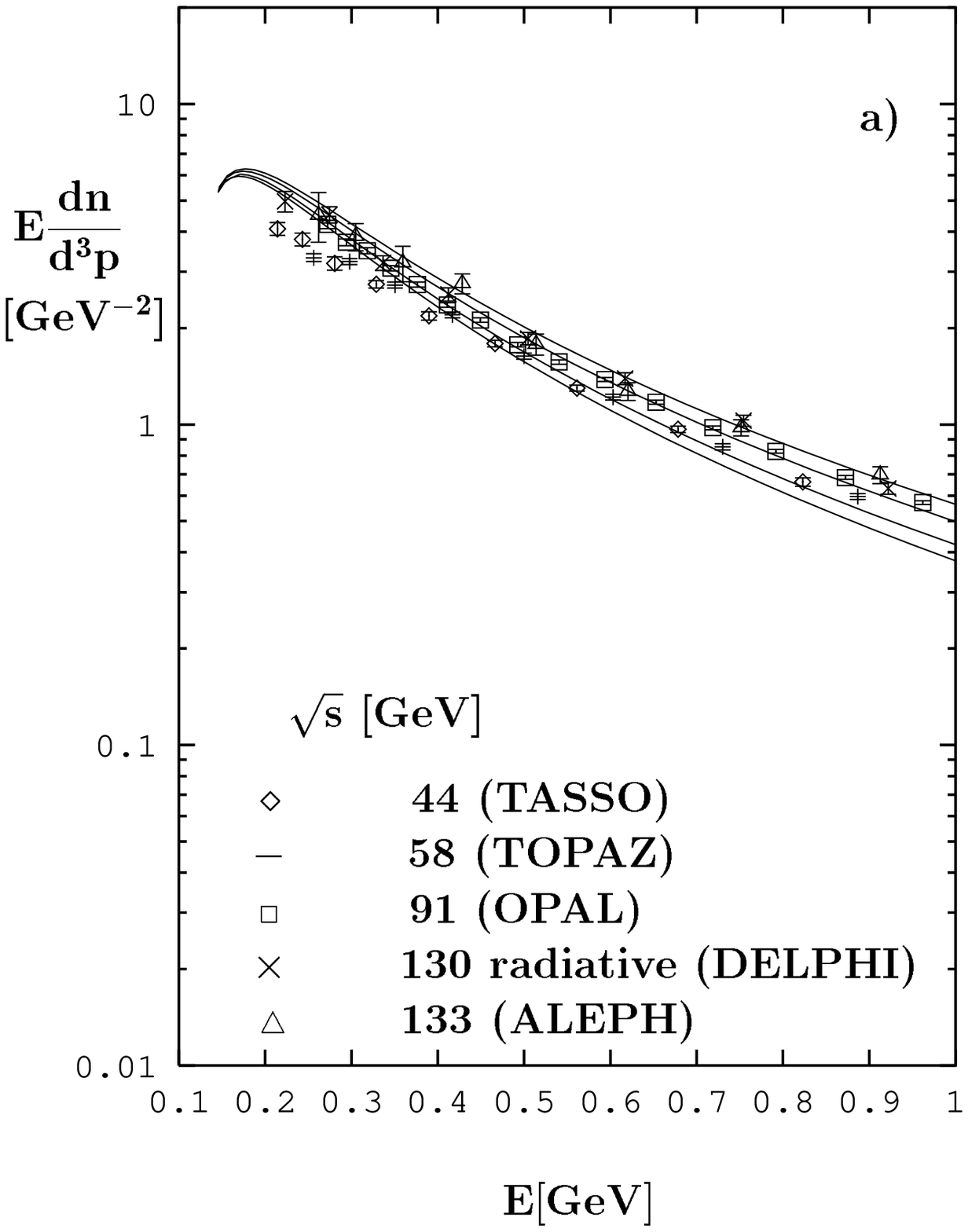,width=.6\linewidth,bbllx=5.5cm,bblly=10.5cm,bburx=13.5cm,bbury=26.5cm}}
          \end{center}
      \end{minipage}\hfill
      \begin{minipage}{.45\linewidth}
          \begin{center}
\mbox{\epsfig{file=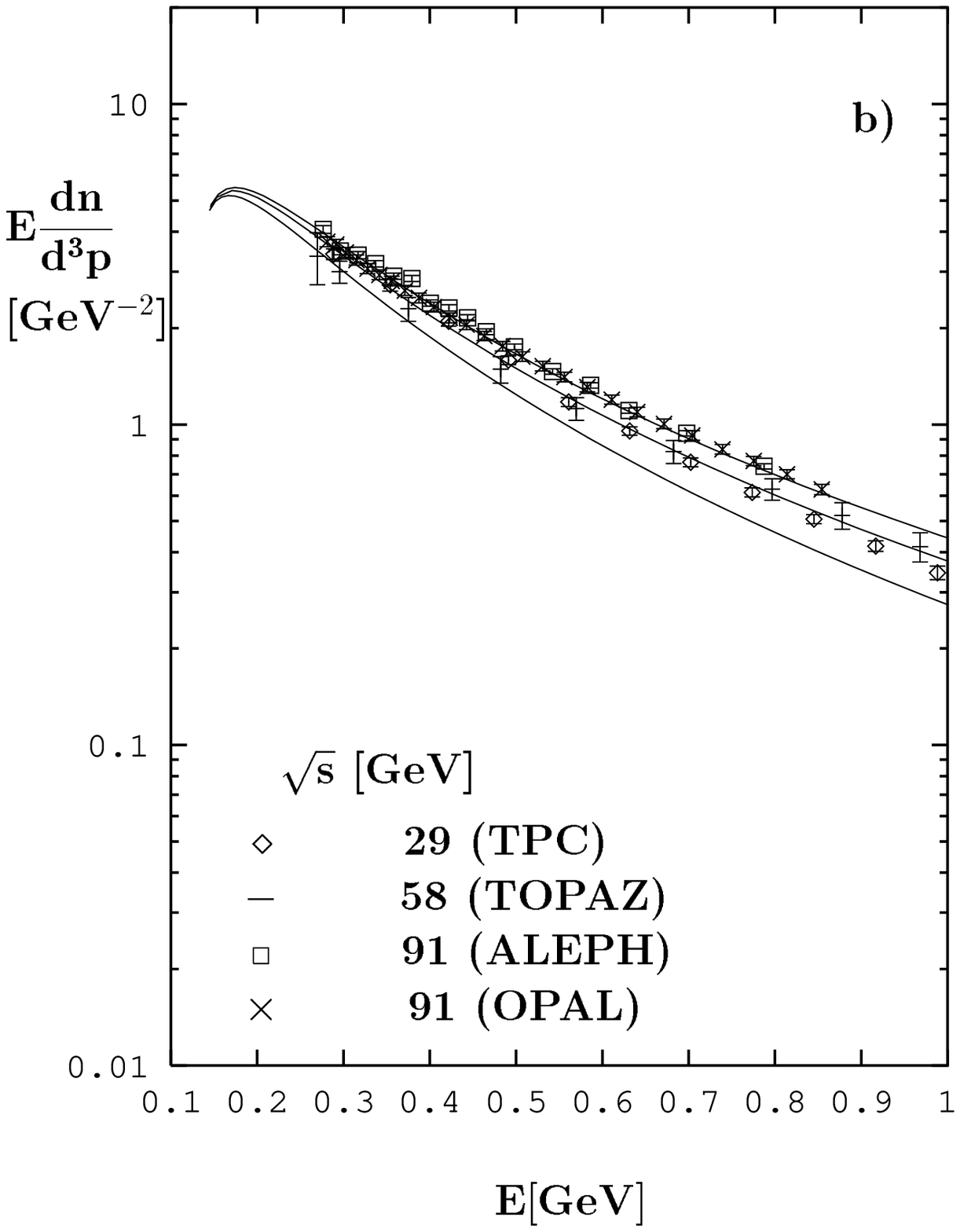,width=.6\linewidth,bbllx=5.5cm,bblly=10.5cm,bburx=13.5cm,bbury=26.5cm}}
          \end{center}
      \end{minipage}
\caption{{\bf a)}: 
invariant density $E dn/d^3p$ of charged particles in $e^+e^-$
annihilation as a function of the particle energy $E = \protect\sqrt{p^2 + m_h^2}$ 
with $m_h$ = 138 MeV at various $cms$ energies compared with the predictions of
eq.~\protect\eref{phrel1} with $Q_0$ = 138 MeV and $K_h$ = 1.125; 
{\bf b)}: same as in {\bf a)}, but for charged pions.} 
\label{khozedata}
\end{figure}

This  approach has been 
 applied both to the invariant density of
all charged particles and to the invariant density of pions. 
Its predictions at different $cms$ energies 
are compared in Fig.~\eref{khozedata} with experimental data 
for charged particles and charged pions. 
At large $cms$ energies, a good agreement is found for charged pions 
in the whole particle energy region and for 
charged particles  in the low particle energy region. 
However, this prescription cannot properly describe data at low $cms$ energies,
i.e., it does not reproduce correctly the $cms$ energy dependence of the data. 

\section{Further tests of LPHD in 3-jet events in $e^+e^-$ annihilation.} 

In view of the success of the LPHD picture for describing the invariant density of
charged particles in $e^+e^-$ annihilation, it is interesting to look for other
independent tests of this picture. 
In this section, a new analysis 
which could shed new light on the validity of the
perturbative description of the soft region is proposed. 

Consider 3-jet events, measure the invariant density of particles 
produced into a cone perpendicular to the production
plane for different angular configurations of the three jets 
and extract then the limiting value $I_0$  for each configuration. 
To get an absolute normalization, 
look also  at the radiation into a cone of the same angle 
 again perpendicular to the primary $q\bar q$ direction, but 
in 2-jet $q\bar q$ events and build the ratio 
\be
R_{\perp} \equiv \frac{dN_{\perp}^{q\bar q g}}{dN_{\perp}^{q\bar q}} 
\ee
The proposed observable seems at a first sight similar to the study of the
string phenomenon, which concerns the radiation in the plane of 3-jet events; 
 the advantage of studying the radiation in the transverse direction 
 lies in the fact that in this case one can avoid 
 the integration over the $k_T \ge Q_0$ boundaries along the jets. 
Moreover, the new observable refers to soft production only, and not to the
total particle flow integrated over momentum, as in the discussion of the 
string effect. For the particle flow, the original angular flow should be
convoluted with a cascading factor\cite{adkt}, which, however, cancels out when
one looks at ratios. The same predictions for the ratio $R_{\perp}$ should then
in principle be valid both for the ratio of the soft particle yields, $I_0$, 
and for the ratio of the multiplicity flows. However, it would be useful to
perform independently both measurements. 

Let us now consider the theoretical prediction for the above observable in the
framework of the perturbative approach. 
The calculation only includes the contribution of 
 soft gluon bremsstrahlung emission up to 
order $\alpha_s$, since, as seen in the study of the invariant density in
$e^+e^-$ annihilation, this term should dominate in the soft limit.

The formula for the soft radiation into arbitrary direction $\vec{n}$
from a $q\bar q$ antenna pointing in directions $\vec{n}_i$ and
$\vec{n}_j$ is given by the usual bremsstrahlung formula\cite{adkt}:
\be
dN_{q\bar q} = \frac{dp}{p} d\Omega_{\vec{n}} 
\frac{\alpha_s}{(2 \pi)^2} W^{q\bar q}(\vec{n}) \quad , \quad 
W^{q\bar q}(\vec{n}) = 2 C_F (\widehat{ij}) 
\label{wqq} 
\ee
with 
$(\widehat{ij}) = a_{ij}/( a_i a_j)$, $a_{ij} = (1 - \vec{n}_i
\vec{n}_j)$ and $a_i = (1 - \vec{n} \vec{n}_i)$. 
For the radiation perpendicular to
the primary partons $(\widehat{ij}) = a_{ij} = 1 - \cos \Theta_{ij}$, with
$\Theta_{ij}$ the relative angle  between the primary partons $i$ and $j$. 

The soft gluon radiation in a 3-jet event is given as in eq.~\eref{wqq} 
with the angular factor
\be
W^{q\bar q g}(\vec{n}) = N_C [ (\widehat{gq}) + (\widehat{g\bar q}) 
- \frac{1}{N_C^2}  (\widehat{q\bar q}) ] 
\ee

The soft gluon radiation for a $q \bar q$ pair with relative angle
$\Theta_{q\bar q}$  is given again as in  eq.~\eref{wqq}, but 
with the angular factor
\be
W^{q\bar q }_{\perp}(\Theta_{q\bar q}) = 2 C_F ( 1 - \cos \Theta_{q\bar q}) \;  .
\ee
which, for 2-jet events in the rest frame, gives 
$W_{\perp}^{q\bar q}(\pi) = 4 C_F$. 

\begin{table}     
 \begin{center}
 \vspace{4mm}
 \begin{tabular}{||c|c|c||}
  \hline
 & $R_{\perp}$ & $R_{\perp}$ (large $N_C$) \\ 
  \hline
 $\Theta_{gq} = \pi - \Theta_{g\bar q}$ & 1 & 1 \\ 
 (collinear or soft gluons) & & \\ 
 $\Theta_{gq} = \Theta_{g\bar q} = \frac{2}{3} \pi$ & 
 1.59  &  
  1.5 \\ 
 (Mercedes) & & \\ 
 $\Theta_{gq} = \Theta_{g\bar q} = \pi$ & $\frac{N_C}{C_F}$ = 2.25 & 2 \\ 
 ($q\bar q$ antiparallel to $g$) & & \\ 
 \hline 
 \end{tabular}
 \end{center}
\caption{Prediction~\protect\eref{rperp}  
for the ratio $R_{\perp} = dN_{\perp}^{q\bar q g}/dN^{q\bar q}_{\perp}$ 
and the large-$N_C$-limit~\eref{rperpln} for different 
angular configurations of the $q\bar q g$ events 
($\Theta_{q\bar q} = 2 \pi - \Theta_{gq} - \Theta_{g\bar q}$).}
\label{tableperp}
\end{table} 

Correspondingly, the ratio $R_{\perp}$ of the soft particle yield in 3-jet events to
that of 2-jet events in their own rest frame  is given by
\be
R_{\perp} \equiv \frac{dN_{\perp}^{q\bar q g}}{dN_{\perp}^{q\bar q}} =
\frac{N_C}{4 C_F} [ 2 - \cos \Theta_{gq} - \cos \Theta_{g\bar q} - \frac{1}{N_C^2}
(1 - \cos \Theta_{q\bar q} ) ] 
\label{rperp}
\ee
Theoretical predictions for $R_{\perp}$ are presented in
Table~\eref{tableperp} for three different angular configurations; notice that
in the two extreme cases, the soft or collinear primary gluon emission 
and the parallel $q \bar q$ configuration, the expected limiting values
are correctly recovered. 
The particularly simple and interesting situation of Mercedes-type events, 
where no jet identification is necessary for the above measurement, is also
shown. 
The table also shows the results obtained in the large-$N_C$ 
approximation, in which the $q\bar q g$ event is treated as a
superposition of two $q\bar q$ dipoles (see, e.g., \cite{lund}). 
In this case the expression for the ratio $R_{\perp}$ simply becomes:
\be
R_{\perp} = 
\frac{1}{2} [ 2 - \cos \Theta_{gq} - \cos \Theta_{g\bar q} ] 
\label{rperpln}
\ee 
The difference among the full theory and the large-$N_C$ limit 
can also be investigated\cite{klo2,nijmegen} 
by studying  the production rate in 3-jet events normalized 
to the sum of rates from the corresponding 2-jet events (dipoles) 
with opening angle $\Theta_{gq}$ and $\Theta_{g\bar q}$ respectively, 
rates which can be extracted experimentally from the analysis of 
$q\bar q \gamma$ events.

\section{Test of universality in different reactions}

Let us look at the soft limit of the invariant density 
\be
I_0 = \lim_{y \to 0, p_T \to 0} E \frac{dn}{d^3p} = 
\frac{1}{2} \lim_{p \to 0} E \frac{dn}{d^3p} 
\ee
in different reactions. 
 This study provides a very direct test of universality in soft particle
 production; one could indeed directly measure whether  the soft
 particle production is universal, i.e., it is a purely hadronic quantity  
 not related to the underlying partonic processes, 
 or the intensity $I_0$ depends  
on the  colour topology  of the primary active partons in the collisions
process, as predicted in our perturbative approach. 
For instance, in case of quark exchange the two outgoing jets
originate from colour triplet charges and $I_0$ should be as in $e^+ e^-$
annihilation; on the contrary, in case of gluon exchange $I_0$ should be 
about twice as large ($N_C/C_F$). 
If data would show these features, 
a  further direct support of the dual description of soft particle
production in terms of the QCD bremsstrahlung would be achieved. 

Unfortunately, an absolute prediction of $I_0$  is not possible, since it 
depends on the normalization factor and the cut-off parameter $Q_0$. 
One should then exploit the energy independence of the soft limit  $I_0$ 
in $e^+e^-$ annihilation and use this value to 
set a  standard scale to be used for the  comparison with other processes. 

\subsection{Quark exchange processes} 

In case the process goes through quark exchange, i.e., via colour triplet
exchange, the soft production intensity $I_0$ should be the same as in
$e^+ e^-$ annihilation.

A possible realization of quark exchange process 
is the process $\gamma \gamma \to q \bar q \to$ 2-jets
with either the virtuality $Q^2$ of the initial photon or
the scattering angle photon-jet sufficiently large (see, e.g.
\cite{kz}).
Another example is deep inelastic scattering at large $Q^2$. In this case
the current fragments in the Breit frame are expected to have the same
characteristics as the quark fragments in one hemisphere of
 $e^+ e^-$ annihilation (for a QCD analysis, see\cite{gdkt}),
and this is indeed observed for not too small $Q^2$~\cite{h1}.
A third case is the production of two W's which decay in the fully hadronic
channel. In this case, since each of the two W's decays into 2 jets, 
the limiting soft particle yield
should be twice the yield in $e^+e^- \to q \bar q$ , 
provided the interconnection phenomena\cite{interconn} are neglected. 
Alternatively, deviations from the factor 2 could give an estimate of the
importance of these effects. 

\subsection{Gluon exchange processes}

In case the process goes through gluon exchange, i.e., via colour octet
exchange, the soft production intensity $I_0$ should be roughly twice as large
($N_C/C_F$) as in $e^+ e^-$ annihilation.
The same factor appears in the theoretical predictions for 
the ratio of the average multiplicity in gluon vs.
quark jets\cite{bg}; experimental analyses of  this ratio 
have obtained values  considerably below this limit (see e.g. \cite{ko}). 
Since in our case we are not integrating over momentum but we are simply
looking at the soft radiation, where some effects, like for instance 
 energy-momentum constraints, are expected to be small, the effects of the 
different colour charges could be more pronounced than in the 
study of the multiplicity ratio. 

The realization of a $gg$ colour singlet final state is not easy 
experimentally.
It may become available at future colliders at higher energies through the
process $\gamma \gamma \to gg$.  An approximate 
realization of a colour octet antenna has been recently obtained
in $e^+e^- \to q \bar q g$, by selecting a configuration 
with the gluon recoiling against a quasi--collinear $q \bar q$ pair\cite{gary} 
(for a recent study of such events, see \cite{opalqq}). 
A process mediated by colour octet exchange is expected to occur also in DIS 
for $Q^2 \gsim$ few GeV$^2$ and small Bjorken $x$, via the diagrams of 
photon gluon fusion. 
Another example is hadron-hadron collision with a particle or jet at
moderate $p_T$ ($\gsim$ 1--2 GeV) at small angles so that the overall 2-jet
structure is maintained. 

\subsection{Soft collisions (minimum bias events)} 

These processes (with initial hadrons or real photons) are not so well
understood theoretically
as the hard ones but it might be plausible to extrapolate the gluon
exchange process towards small $p_T$\cite{ln}.
Accordingly, the soft radiation is expected to be twice as large 
as in $e^+e^-$ annihilation. 
For these reactions, some experimental information is already available. 
Data at ISR energies show that the soft limit $I_0$ 
is similar to $e^+ e^-$ annihilation at a similar effective energy\cite{ppscal}. 
On the other hand, $I_0$ roughly doubles when going from $\sqrt{s} \sim$ 20 GeV
\cite{na22,ISR} to $\sqrt{s}$ = 900~GeV  at the collider\cite{coll}.
If additional incoherent sources  can be excluded, such behaviour
could indicate the growing importance of one-gluon exchange expected from the
perturbative picture. Then a saturation at $I_0^{hh}/I_0^{e^+e^-} \sim 2$
 and no additional increase for a semihard $p_T$ trigger would be expected.
However, the rise of $I_0^{hh}$ at collider energies could also result 
from the incoherent multiple collisions of partons (e.g. \cite{pythia}) 
which has recently  been postulated in $\gamma p$ collisions\cite{sz}. 
This unclear situation certainly deserves further studies. 

\subsection{Rapidity dependence of $I_0$ in DIS} 

According to the previous discussion, different kinematical conditions 
in $\gamma p$ collisions select different elementary subprocesses and should
then give rise to different yields $I_0$'s. 
For example, by decreasing $Q$ at fixed hadronic energy $W$ 
one goes from quark exchange process to a gluon exchange process and finally to
a soft process. We have seen that different reference 
frames, for instance the Breit
frame,  can be useful to select a particular situation. 
An alternative possibility is provided by the study of 
the soft particle yield $I_0$ at different values of rapidity in 
one reference frame, for instance the cms frame. 
Let us consider then the yield $I_0(y)= \frac{dn}{dy d^2p_T} 
\bigr|_{p_T \to 0}$ as a function of rapidity $y$, 
measured in the $cms$ frame. 
One would then expect
the existence of a ``quark plateau" of
$I_0(y)$ in the current region of length $\Delta y \approx \log (Q/m)$ and
height equal to $e^+e^-$ annihilation 
near the rapidity corresponding to the Breit frame, i.e., in the direction of
the incoming photon, $y_{Breit}$; moving towards the central region, 
 a transition to a ``gluon plateau" in the complementary region is expected to
 occur. If sufficiently hard gluons are exchanged 
 in the process, $I_0$ should then increase  but still remain 
smaller than twice the value in $e^+e^-$ annihilation. 
Approaching the fragmentation region, a transition to soft interactions should
be observed and a decreasing of $I_0$ is likely to occur. 
In general, the soft particle yield $I_0$ as a function of the rapidity should
show a step like behaviour, pointing out the regions where the underlying
process is dominated by quark or gluon exchange; a direct indicator of the
underlying mechanism would be provided directly by the height of the plateau. 

\section{Conclusions}

The analytical perturbative approach to multiparticle production, based on the
Modified Leading Log Approximation and Local Parton Hadron Duality 
has been shown to successfully describe  data on inclusive energy
spectra of charged particles in QCD jets. In order to investigate the
limitations of this picture and to point out non perturbative effects, this
approach has been applied in the soft region. 

The invariant densities $E dn/d^3p$ of all 
charged particles and identified particles in $e^+e^-$ annihilation 
are found to be 
approximately energy independent at low particle energy over a very large
range of $cms$ energies. 
The same behaviour is expected in the perturbative approach, as a consequence
of the coherence of the soft gluon radiation from all emitters. 
With proper additional assumptions for the treatment of mass effects, the 
perturbative predictions reproduce the data also quantitatively, thus 
extending the phenomenological success of the perturbative approach down to the
soft region. This result suggests that  the production of hadrons in the soft
region, which is known 
to proceed through many resonance channels, can be simply parametrized 
through a parton cascade pushed down to a small scale of a few 100 MeV. 

The study of transverse radiation in multijet events in $e^+e^-$ annihilation 
has been proposed to test the sensitivity of the soft particle production 
to the effective colour  charge of the primary emitters. 
Predictions of the perturbative approach for the soft particle production 
in different reactions have also been presented. The perturbative approach 
 predicts a breakdown of universality among different reactions, since
the soft particle yield reflects  in this picture the properties of the
underlying partonic process. 

\section*{Acknowledgements} 
I thank the Directors of the Workshop, L. Cifarelli, 
A. Kaidalov and V. A. Khoze, for the invitation and the nice atmosphere
created at this Workshop. I thank Wolfgang Ochs and Valery A. Khoze for
discussions and collaboration on the subject of this talk.

\end{document}